\documentclass[reprint,amsmath,amssymb,aps,prb]{revtex4-2}

\usepackage{graphicx}
\usepackage{dcolumn}
\usepackage{bm}
\usepackage{layouts}
\usepackage{verbatim}
\usepackage[dvipsnames]{xcolor}
\usepackage{siunitx}
\usepackage{placeins}

\begin{document}
\title{Electrical detectability of magnon-mediated spin current shot noise}

\author{Luise Siegl$^1$}
\author{Michaela Lammel$^1$}
\author{Akashdeep Kamra$^2$}
\author{Hans Huebl$^{3,4,5}$}
\author{Wolfgang Belzig$^1$}
\author{Sebastian T. B. Goennenwein$^1$}
\affiliation{$^1$Department of Physics, University of Konstanz, 78457 Konstanz, Germany \\
			 $^2$Condensed Matter Physics Center (IFIMAC) and Departamento de Física Teórica de la Materia Condensada, Universidad Autónoma de Madrid, E-28049 Madrid, Spain \\
			 $^3$Walther-Meißner-Institut, Bayerische Akademie der Wissenschaften, 85748 Garching, Germany\\
			 $^4$TUM School of Natural Sciences, Technische Universität München, 85748 Garching, Germany\\
			 $^5$Munich Center for Quantum Science and Technology (MCQST), 80799 München, Germany
			}

\date{\today}

\begin{abstract}
A magnonic spin current crossing a ferromagnet-metal interface is accompanied by spin current shot noise arising from the discrete quanta of spin carried by magnons. 
In thin films, e.g., the spin of so-called squeezed magnons have been shown to deviate from the common value $\hbar$, with corresponding changes in the spin noise. 
In experiments, spin currents are typically converted to charge currents via the inverse spin Hall effect.
We here analyze the magnitude of the spin current shot noise in the charge channel for a typical electrically detected spin pumping experiment, and find that the voltage noise originating from the spin current shot noise is much smaller than the inevitable Johnson-Nyquist noise.
Furthermore, we find that due to the local nature of the spin-charge conversion, the ratio of spin current shot noise and Johnson-Nyquist noise cannot be systematically enhanced by tuning the sample geometry, in contrast to the linear increase in dc spin pumping voltage with sample length. 
Instead, the ratio depends sensitively on material-specific transport properties.
Our analysis thus provides guidance for the experimental detection of squeezed magnons through spin pumping shot noise.
\end{abstract}

\maketitle

The power spectral density of charge current fluctuations contains fundamental information about the underlying transport and dynamics~\cite{Blanter2000}. 
For example, the discrete nature of electric charge results in shot noise~\cite{Schottky1918}. 
Vice-versa, shot noise experiments allow to quantify the quantum of charge relevant for transport. 
In diodes and related structures, the existence of shot noise shows that the electrical current is carried by elementary charges, while in fractional quantum Hall systems, composite Fermions with fractional charge are the relevant electrical transport quanta~\cite{de-Picciotto1998,Saminadayar1997,Ferrier2016}. 
In superconducting contacts, multiple charge quanta have been predicted~\cite{Khlus1987,Cuevas1999,Naveh1999,Cuevas2003} and observed~\cite{Jehl2000,Kozhevnikov2000,Cron2001}.
On the other hand, thermal fluctuations of charge carriers inside an electrical conductor at equilibrium lead to the Johnson-Nyquist noise~\cite{Johnson1928,Nyquist1928}.

In recent years, pure spin transport has attracted considerable interest. 
In particular, spin pumping has emerged as a powerful method for the generation of pure spin currents in ferromagnetic/normal metal (FM/N) heterostructures~\cite{Urban2001, Tserkovnyak2002a, Tserkovnyak2002b, Azevedo2005, Saitoh2006, Kajiwara2010, Wang2014, Liu2020}. 
There, a magnon mode in the FM is populated using a coherent microwave drive, i.e., in ferromagnetic resonance (FMR). 
The resulting nonequilibrium magnonic spin is partially absorbed by the electrons in N causing a pure spin current flow across the FM/N interface. 
Taking advantage of the (inverse) spin Hall effect which interconverts pure spin and charge currents in a metal with strong spin-orbit coupling~\cite{Hirsch1999, Zhang2000,Sinova2015}, such spin currents can be detected in N as an electrical current or voltage signal.

Thermal fluctuations of a pure spin current have been detected experimentally in a yttrium iron garnet/platinum bilayer employing a magnetic field orientation-dependent measurement of the voltage noise power spectral density~\cite{Kamra2014}.
The observed thermal spin current noise was theoretically shown to be related to the spin Hall magnetoresistance (SMR) effect via the fluctuation-dissipation theorem~\cite{Kamra2014}.
The SMR effect is governed by spin current flow across a  magnetic insulator/metal interface~\cite{Chen2013, Nakayama2013, Vlietstra2013a, Althammer2013, Weiler2013}. 
More precisely, the resistance of the metal layer changes as a function of the magnetization orientation due to spin current flow across the interface.
However, while serving as a proof-of-concept for spin current noise measurements~\cite{Kamra2014}, the observed thermal voltage noise in platinum did not provide deeper insights into the microscopic mechanisms of spin transport.

The situation is different for fluctuations of a non-equilibrium current~\cite{Brataas2000, Belzig2004}, such as spin shot noise arising from a pure spin current $I_\mathrm{s}$ flowing across the FM/N interface~\cite{Kamra2016a, Kamra2016b}, or non-equilibrium spin accumulations~\cite{Meair2011}. 
In analogy to electrical shot noise, this spin current shot noise can be used to experimentally detect and quantify the spin transport quantum~\cite{Kamra2016a,Kamra2016b}. 
For spin transport arising from a coherently driven magnon mode, as is the case in a typical ferromagnetic resonance scenario, the non-integer character of the magnon spin transport quantum, e.g. due to squeezing effects, could become experimentally accessible.
The same information is harder to infer from a thermally driven spin current shot noise, for example via the spin Seebeck effect~\cite{Matsuo2018}, as it involves multiple magnon modes with different effective spins.

However, previous theoretical works~\cite{Kamra2016a,Kamra2016b} on spin current shot noise have restricted themselves to examine spin currents, with only a preliminary discussion on spin to charge current noise conversion~\cite{Matsuo2018,Joshi2018,Aftergood2020}.
Moreover, to assess the experimental detectability, the additional noise contributions arising from charge fluctuations in the normal metal must be taken into account. 
Here, we calculate the magnitude of the spin current shot noise power spectral density upon conversion to the charge channel, and consider how the spin to charge conversion impacts the detectability of spin current shot noise. 
A key prerequisite for this consideration is that the spatial correlations in the electronic spin current injected into N are short ranged comparable to the electronic wavelengths. 
This implies that the conversion factor for spin to charge dc currents, widely used in spin-Hall effect based studies, does not provide a complete picture for the conversion of the interfacial spin current to the electrically measured voltage fluctuations. 
In consequence, we find that the voltage noise resulting from the spin current shot noise is substantially smaller than the purely charge based Johnson-Nyquist (JN) noise in the normal metal. 

We consider the shot noise associated with the spin pumping current in a system at a finite temperature $T$, driven by a coherent microwave magnetic field at driving frequency $\omega$~\cite{Kamra2016b} which corresponds to the ferromagnetic resonance frequency. 
The ensuing power spectral density in the spin channel for a total spin current $I_{\mathrm{s}}$ traversing the FM/N interface is given by $S_{I_\mathrm{s}I_\mathrm{s}} = 2\hbar^*I_\mathrm{s}\frac{2k_{\rm{B}}T}{\hbar\omega}$~\cite{Kamra2016b}, where $k_{\rm{B}}$ is the Boltzmann constant, $\hbar$ is the reduced Planck constant and $\hbar^*$ the effective spin. 
Interestingly, spin shot noise thus reflects magnon squeezing effects, since the quantum of angular momentum relevant for the noise is no longer $\hbar$, but $\hbar^* = \hbar (1+\delta)$~\cite{Kamra2016a}, where $\delta$ is a material and sample geometry dependent factor. 
These results suggest that spin current shot noise can be used to experimentally detect and quantify magnon squeezing by the effective spin $\hbar^*$ via spin pumping experiments.

\begin{figure}
	\includegraphics{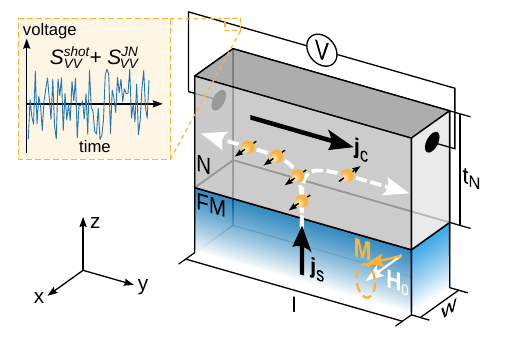}%
	\caption{Schematic of a typical spin pumping experiment with electrical detection via the inverse spin Hall effect. The resonantly driven coherent precession of the magnetization $\bm{M}$ in the ferromagnetic layer (FM) drives a pure spin current density $\bm{j}_{\rm{s}}$ propagating in the z-direction, across the interface into the  normal metal (N) layer. In the latter, it is converted into a charge current density $\bm{j}_{\rm{c}}$ owing to the ISHE. For open circuit electrical boundary conditions, a dc voltage $V$ proportional to $\bm{j}_{\rm{s}}$ arises. \label{fig1:ISHE}}
\end{figure}

Now, we consider the experimental detection scheme to measure spin current shot noise electrically via the inverse spin Hall effect (ISHE) in the N layer. 
Within the FM layer the magnetisation $\bm{M}$ is coherently driven out of its equilibrium position by FMR~\cite{Griffiths1946, Kittel1947, Kittel1948}. 
Thus, a pure spin current $I_\mathrm{s} = j_{\rm{s}}wl$ propagating along z-direction is pumped over the interface of width $w$ and length $l$ into the metal layer, as depicted in Fig.\,\ref{fig1:ISHE}. 
Note that for this expression of the total spin pumping current, we have assumed a spatially uniform spin pumping current density $j_{\rm{s}}$, which is valid for a dc current or the expectation value of an ac current.
Since a direct experimental detection of spin is difficult, the spin current density is converted into a charge current density $\bm{j}_{\rm{c}} = \frac{2e}{\hbar} \theta_{\rm{SH}}\,\bm{j}_{\rm{s}} \times \bm{s}$  via the inverse spin Hall effect in N~\cite{Hirsch1999, Saitoh2006, Hoffmann2013, Sinova2015}, as sketched in Fig.\,\ref{fig1:ISHE}. Here, $e$ is the elementary charge, and $\theta_{\rm{SH}}$ the spin Hall angle. 
In most experiments, open circuit electrical boundary conditions are implemented, such that an open-circuit dc voltage $V$ is detected instead of the charge current.

Applying such an ISHE-based electrical detection scheme to spin current shot noise, one obtains a voltage noise power spectral density $S_{VV}\propto S_{I_{\rm{s}}I_{\rm{s}}}$ in the charge channel inside the normal metal. 
Generally, the spin shot noise can be detected as an electrical voltage noise power spectral density $S_{VV}$ or current noise power spectral density $S_{II}$.
 Since the current and voltage power spectral densities can be transformed into each other by $\left(\frac{S_{VV}}{V^2}\right)_{I=\rm{const}} = \left(\frac{S_{II}}{I^2}\right)_{U=\rm{const}}$~\cite{Kogan1996}, we here focus on $S_{VV}$.

The typical theoretical analysis~\cite{Mosendz2010} exploits the spatial homogeneity of the spin pumping current density $\bm{j}_{\rm{s}}$ over the FM/N interface to relate $\bm{j}_{\rm{s}}$ with the experimentally measurable quantity, i.e. the total charge current $I$ (or voltage $V$) through the normal metal.
This, in turn, is a consequence of the spatial invariance of the coherent microwave drive causing FMR and thus, the magnetization precession in the FM layer. 
On the other hand, the spin pumping current noise or fluctuations are expected to have short ranged correlations determined by the wavelength of the electrons that absorb and carry the spin current in N. 
Thus, we need to go beyond the typical relation for dc currents~\cite{Mosendz2010,Matsuo2018}, as discussed below, to relate the power spectral density of the total spin pumping current noise $S_{I_\mathrm{s}I_\mathrm{s}}$~\cite{Kamra2016a,Kamra2016b} to the total charge current $I$ through the normal metal.
The spin current shot noise $S_{I_\mathrm{s}I_\mathrm{s}}$ scales with the system temperature and is largest under the condition of ferromagnetic resonance $\omega=\omega_0$~\cite{Kamra2016b}. 
We thus exploit the result from Ref.~\cite{Kamra2016b} in the high temperature limit $k_{\rm{B}}T \gg \hbar\omega$ for sufficiently low driving frequencies $\omega$, as it is the experimentally relevant limit. 
The low temperature limit $k_{\rm{B}}T \ll \hbar\omega$ is briefly discussed below.
Assuming a y-polarized spin current density $j_{\rm{s}}$, we obtain the spatially and temporally resolved spin pumping current density correlator 
\begin{equation}\label{eq:corr_func}
	\langle j_{\rm{s}}(t,\bm{\rho})j_{\rm{s}}(t',\bm{\rho'}) \rangle = 2 \hbar^* j_{\rm{s}} \frac{2k_{\rm{B}} T}{\hbar\omega_0} \delta (t-t') \delta (\bm{\rho}-\bm{\rho'})
\end{equation}
local in time $t$ and space, where $\bm{\rho}$ is the two-dimensional position vector in the interfacial plane.
Equation \eqref{eq:corr_func} captures the low-frequency and frequency-independent part of the spin current noise power spectral density. 
It has been derived starting from the correlator of the total spin current across the interface~\cite{Kamra2016b}, and considering a coherent region. 
Assuming additionally that the coherence length is much smaller than the sample dimensions, a delta function in space is obtained.
Taking this spatio-temporal correlation for the interfacial spin current density and following an analysis similar to that in Ref.~\cite{Kamra2014}, we evaluate the voltage noise power spectral density $S_{VV}^{\rm{shot}}$ of the spin pumping current shot noise in the charge channel
\begin{equation}\label{eq:SVS(js)}
	S_{VV}^{\rm{shot}} = 16 \frac{\theta_{\rm{SH}}^2 \lambda_{\rm{sd}}^2 \rho_{\rm{N}}^2 l}{w t_{\rm{N}}^2} j_{\rm{s}} e^2 \frac{\hbar^*}{\hbar^2} \frac{2k_{\rm{B}} T}{\hbar\omega_0} \tanh^2\left(\frac{t_{\rm{N}}}{2 \lambda_{\rm{sd}}}\right) .
\end{equation}
Here, $\omega_0$ is the FMR frequency, and $\lambda_{\rm{sd}}$, $\rho_{\rm{N}}$ and $t_{\rm{N}}$ are the spin diffusion length, the resistivity and the thickness of the metal layer N, respectively.

We now interpret the shot noise enhancement with temperature~\cite{Kamra2016b}. 
We begin by recognizing that the dc spin current expression contains an integral of the form $\sim\int d\epsilon [f(\epsilon) - f(\epsilon + \hbar\omega)]$, where $\epsilon$ denotes energy, $\omega$ is the magnon (microwave drive) frequency, and $f(\epsilon)$ is the Fermi function~\cite{Kamra2016b}. 
The physical content is that the rate of spin transfer is determined by the difference between available electrons and holes (unoccupied electronic states) within $\hbar\omega$. 
This is reasonable since a magnon is absorbed by raising an electron from an occupied to an unoccupied state or by creating a pair of electron and hole excitations. 
On the other hand, the expression for the shot noise contains an integral $\sim\int d\epsilon f(\epsilon) [1 - f(\epsilon + \hbar\omega)]$ which shows that the coherently driven and populated magnon mode is causing correlations between existing electron and hole excitations~\cite{Kamra2016b}. 
At zero temperature, since there are no pre-existing excitations, the only way to create a correlation is by exciting them at the same time via the absorption of a magnon. 
However, at finite temperatures, the macroscopically occupied magnon mode may induce correlations in the thermally excited electron-hole pairs resulting in stronger correlations that are still proportional to the number and spin of magnons. 
In mathematical terms, at zero temperature, the integrands in both the integrals above are unity and nonzero only over a $\hbar\omega$-wide energy window resulting in their proportionality. 
While finite temperatures do not affect the integral for the dc current, they significantly enhance the window over which electron and hole excitations coexist and the integrand is nonzero. 
The shot noise induced correlations are thus enhanced with temperature. 
To conclude, the simple picture that shot noise results solely from the quantized nature of transport is only valid at zero temperature.

In addition to the voltage noise, Eq.~\eqref{eq:SVS(js)}, arising from the spin pumping current traversing the FM/N interface, the normal metal harbors a thermal charge fluctuations-based Johnson-Nyquist (JN) noise with a power density 
\begin{equation}\label{eq:S_JN}
	S_{VV}^{\rm{JN}} = 4 k_{\rm{B}} T R.
\end{equation}
Here, $R = \rho_{\rm{N}} l/(t_{\rm{N}} w)$ is the resistance of N. 
Note, that the noise represented by Eq.~\eqref{eq:S_JN} is different from the contribution of the thermal magnonic spin current fluctuations~\cite{Kamra2016b,Matsuo2018}, which can be considered a magnonic spin transport analogue of the JN noise. 
As the thermal magnonic spin current fluctuations has been theoretically shown to be smaller than the shot noise in a wide range of parameters~\cite{Kamra2016b,Matsuo2018} we disregard this magnonic contribution in our analysis here.
The voltage noise in N thus will have (at least) two contributions, $S_{VV} = S_{VV}^{\rm{shot}} + S_{VV}^{\rm{JN}}$. 
Since both are frequency independent (white) at low frequencies, we consider and compare their absolute magnitudes.
The ratio 
\begin{equation}\label{eq:SVS/SJN}
	\frac{S_{VV}^{\rm{shot}}}{S_{VV}^{\rm{JN}}} = 8 \frac{\theta^2_{\rm{SH}} \lambda^2_{\rm{sd}} \rho_{\rm{N}}}{t_{\rm{N}}} j_{\rm{s}} \frac{\hbar^*e^2}{\hbar^3\omega_0} \tanh^2\left(\frac{t_{\rm{N}}}{2 \lambda_{\rm{sd}}}\right) 
\end{equation}
should be maximized for the detection of spin current shot noise. 
Since $\delta$ is on the order of $1$ in thin ferromagnetic films, we use $\hbar^*=2\hbar$ to estimate the ratio~\cite{Kamra2016a}. 
Therefore, the adjustable parameters for maximizing the spin noise voltage signal are the magnitude of the spin current density $j_{\rm{s}}$, the FMR frequency $\omega_0$, and the thickness $t_{\rm{N}}$ of the metal layer together with its material-specific properties $\theta_{\rm{SH}}$, $\lambda_{\rm{sd}}$ and $\rho_{\rm{N}}$. 
Similar conclusions have been drawn by Luo et al.~regarding the magnitude of magnon shot noise in spin Seebeck effect measurements~\cite{Luo2023}.

In finding the optimal sample design for electrical spin current shot noise experiments, we first note that only the ratio of spin diffusion length $\lambda_{\rm{sd}}$ and metal layer thickness $t_{\rm{N}}$ enters in Eq.~\eqref{eq:SVS/SJN}. 
Since $t_{\rm{N}}$ can be straightforwardly chosen by appropriate sample design, we numerically optimize the expression
\begin{equation}\label{eq:f(lambda,tN)}
	p=\frac{\lambda_{\rm{sd}}}{t_{\rm{N}}}\tanh^2\left(\frac{t_{\rm{N}}}{2 \lambda_{\rm{sd}}}\right)
\end{equation}
from Eq.~\eqref{eq:SVS/SJN}.
Using Newton's method we find numerically that Eq.~\eqref{eq:f(lambda,tN)} has a global maximum at $p\approx 0.29$ for $t_{\rm{N}}\approx 2.18\lambda_{\rm{sd}}$.

\begin{figure}[t]
	\includegraphics{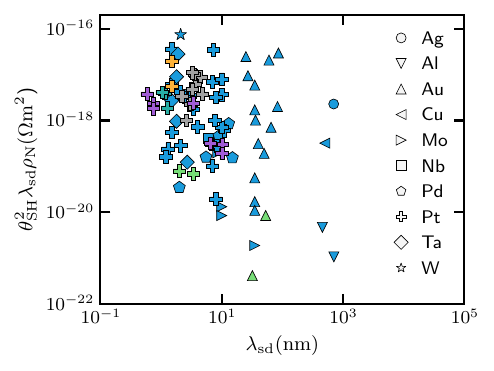}%
	\caption{The combination of material parameters $\theta^2_{\rm{SH}}\lambda_{\rm{sd}}\rho_{\rm{N}}$ relevant for  Eq.~\eqref{eq:SVS/SJN} for different normal metals N. The data were taken from Sinova et al.~\cite{Sinova2015} (blue), Hong et al.~\cite{Hong2018} (grey), Ma et al.~\cite{Ma2018a} (dark green), Isasa et al.~\cite{Isasa2015} (light green), Sagasta et al.~\cite{Sagasta2016} (purple) and Weiler et al.~\cite{Weiler2013} (yellow). The symbols correspond to different materials.
		\label{fig2:theta2lambda1rho-vs-lambda}}
\end{figure}

Next, we consider the material properties of the normal metal N in the combination  $\theta^2_{\rm{SH}}\lambda_{\rm{sd}}\rho_{\rm{N}}$ in Eq.~\eqref{eq:SVS/SJN}. 
Figure~\ref{fig2:theta2lambda1rho-vs-lambda} shows a compilation of values from literature for a set of different spin Hall active metals~\cite{Sinova2015, Hong2018, Ma2018a, Isasa2015, Sagasta2016, Weiler2013}.  
The symbols hereby indicate different materials, the colors the references from which the  values were taken.  
As evident from Fig.~\ref{fig2:theta2lambda1rho-vs-lambda}, the values of $\theta^2_{\rm{SH}} \lambda_{\rm{sd}} \rho_{\rm{N}}$ span 5 orders of magnitude.
This large variation reflects the broad scatter in spin Hall angles and spin diffusion lengths reported even for the same material~\cite{Sinova2015, Hong2018, Ma2018a, Isasa2015, Sagasta2016, Weiler2013}. 

Finally, we turn to the magnitude of the spin current  density $j_{\rm{s}}$. 
In spin pumping experiments, values of $\SI{1.4e-11}{J/m^2} < j_{\rm{s}} < \SI{8.8e-9}{J/m^2}$ have been reported~\cite{Weiler2013,Nakayama2012}. 
Note that Weiler et al. found that the spin Seebeck effect allows generating larger $j_{\rm{s}}$ than those obtained from spin pumping~\cite{Weiler2013}, presumably due to the contribution of magnons with a broad range of wave vectors and frequencies. However, the non-integer effective spin $\hbar^*$ of the squeezed magnon is largest for the $\bm{k}=0$ eigenmode (Kittel mode) and $\hbar^* \to \hbar$ with increasing $\bm{k}$~\cite{Kamra2020}.
Therefore, experiments based on the spin Seebeck effect appear less suitable for investigating the basic mechanisms behind spin current shot noise and magnon squeezing effects. We therefore here focus on pure spin currents generated via spin pumping.

\begin{figure}[b]
    \includegraphics{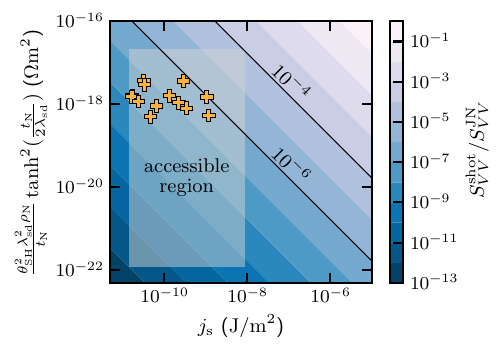}%
    \caption{False color plot showing the voltage noise power density due to spin current shot noise in relation to the Johnson-Nyquist voltage noise density, $S_{VV}^{\rm{shot}}/S_{VV}^{\rm{JN}}$, as given by Eq.~\eqref{eq:SVS/SJN} for a FMR frequency $\omega_0/2\pi=\SI{10}{GHz}$ and $\hbar^*=2\hbar$. The semi-transparent rectangle  highlights the accessible range of $S_{VV}^{\rm{shot}}/S_{VV}^{\rm{JN}}$ values considering literature values~\cite{Sinova2015, Hong2018, Ma2018a, Isasa2015, Sagasta2016, Weiler2013, Nakayama2012} for the spin current density $j_{\rm{s}}$ achievable in spin pumping experiments and the combination of spin Hall angle $\theta_\mathrm{SH}$, spin diffusion length $\lambda_\mathrm{sd}$ and resistivity $\rho_\mathrm{N}$ for different spin Hall active normal metals of thickness $t_\mathrm{N}$ (see main text). The yellow symbols correspond to the experimental data taken from Weiler et al.~\cite{Weiler2013}.
    \label{fig4:SspinSJN_vs_thetalambdarho_vs_js}}
\end{figure}

Taking together the previous results we can extract the ratio of the spin current shot noise and Johnson-Nyquist noise for different parameter combinations.
Figure~\ref{fig4:SspinSJN_vs_thetalambdarho_vs_js} shows this ratio in the voltage channel, $S_{VV}^{\rm{shot}}/S_{VV}^{\rm{JN}}$, as given by Eq.~\eqref{eq:SVS/SJN}. 
We hereby assumed a FMR frequency of $\omega_0/2\pi=\SI{10}{GHz}$ as typical for measurements with a \mbox{X band} microwave cavity, and $\hbar^*=2\hbar$ as mentioned above. 
Based on the parameter values discussed in the preceding paragraphs, the range of experimentally achievable noise power ratios is indicated as a semi-transparent rectangle in the figure.
Notably, even for the best possible combination of material parameters and spin current densities reported in the literature the ratio $S_{VV}^{\rm{shot}}/S_{VV}^{\rm{JN}}$ is smaller than $10^{-4}$. 
This upper experimental boundary is marked by the black line in Fig.~\ref{fig4:SspinSJN_vs_thetalambdarho_vs_js}. 
In previous experiments addressing the voltage noise due to thermal spin current fluctuations, changes in the noise magnitude of $\approx\num{1e-3}$ of the Johnson-Nyquist noise~\cite{Kamra2014} could be resolved. 
We furthermore include data from Weiler et al. (yellow symbols)~\cite{Weiler2013} in Fig.~\ref{fig4:SspinSJN_vs_thetalambdarho_vs_js}, as a typical example for electrically detected spin pumping data recorded using yttrium iron garnet/Pt thin film bilayers and a \mbox{X band} microwave cavity. 
In these experiments, the ratio of the spin current shot noise and Johnson-Nyquist noise is smaller than $10^{-6}$. 
Note that, while in cavity-based FMR a frequency of $\SI{10}{GHz}$ is widely used, it would be beneficial for spin current shot noise experiments to reduce the microwave frequency as much as possible to increase the shot noise.
In the low temperature limit where $k_{\rm{B}}T \ll \hbar\omega$, the Johnson-Nyquist voltage noise power spectral density is suppressed exponentially\cite{Callen1951,Kubo1966}.
However, in this regime other noise sources, e.g. originating from quantum fluctuations might dominate~\cite{Clerk2010, Mariantoni2010}. 
This warrants a quantitative analysis of fluctuations associated with quantum effects, which is beyond the scope of this work.

Our analysis thus indicates that for the detection of the spin current shot noise in an electrical experiment, both the sample properties and the spin current drive need to be carefully optimized.  
Furthermore, to resolve the effective spin $\hbar^*$ of the magnon from spin pumping driven experiments~\cite{Kamra2016a}, even higher experimental precision will be required. 

In summary, we derived the correlator of the spin pumping current density including its time and spatial dependence. 
Considering the spin-to-charge conversion process typically used in electrically detected spin current experiments, we find that in the voltage channel, the spin current shot noise is small compared to the ubiquitous Johnson-Nyquist noise.
More precisely, the ratio of the spin current shot noise to Johnson-Nyquist noise is estimated to be at most $10^{-3}$ using parameters from literature and assuming a driving frequency of $\SI{1}{GHz}$.
We thus conclude that a careful choice of materials is of key importance for the measurement of the spin pumping current shot noise and thus the effective spin of squeezed magnon.
Hence, our work offers important guidance regarding sample design and optimization for the experimental detection of spin current shot noise.

\FloatBarrier

\begin{acknowledgments}
We acknowledge financial support from the Deutsche Forschungsgemeinschaft (DFG, German Research Foundation) via the SFB 1432 – Project-ID 425217212. 
\mbox{H. Huebl} acknowledges financial support by the Deutsche Forschungsgemeinschaft (DFG, German Research Foundation) via Germany’s Excellence Strategy EXC-2111-390814868. 
A. Kamra acknowledges financial support from the Spanish Ministry for Science and InnovationAEI Grant CEX2018-000805-M (through the “Maria de Maeztu” Programme for Units of Excellence in R\&D) and grant RYC2021-031063-I funded by MCIN/AEI/10.13039/501100011033 and “European Union Next Generation EU/PRTR”.
\end{acknowledgments}

\bibliography{2023-09-13_references}

\end{document}